\documentclass[journal]{IEEEtran}
\usepackage{tikz}
\usetikzlibrary{positioning, arrows.meta, fit, calc, shapes.geometric, backgrounds}
\usepackage{cite}
\usepackage{amsmath,amssymb,amsfonts}
\usepackage{algorithm}
\usepackage{algorithmic}
\usepackage{graphicx}
\usepackage{textcomp}
\usepackage{xcolor}
\usepackage{bm} 
\usepackage{booktabs} 
\usepackage{caption}

\usepackage{hyperref}

\begin{document}
\captionsetup[figure]{labelfont={rm},labelformat={default},labelsep=period,name={Fig.}}
\title{Heterogeneous Mixture-of-Experts for Energy-Efficient Multimodal ISAC in Highly Mobile Networks}

\author{Wenqi Fan,~Ning Wei,~\IEEEmembership{Member,~IEEE},~Rongyan Xi,~Ahmad Bazzi,Yue Xiu,~\IEEEmembership{Member,~IEEE},\\
~Chadi Assi,~\IEEEmembership{Fellow,~IEEE},~Jing Dong,~Jing Jin
\thanks{Wenqi Fan, Ning Wei and Yue Xiu are with 
National Key Laboratory of Science and Technology on Communications, University of Electronic Science and Technology of China, Chengdu 611731, China (E-mail:  
fwq897032532@gmail.com, wn@uestc.edu.cn, xiuyue12345678@163.com).
Chadi Assi is with Concordia Institute for Information Systems Engineering,
Concordia University, Quebec H3G 1M8, Canada (e-mail: Assi@ciise.concordia.ca).
Ahmad Bazzi is with the Wireless Research Lab, New York University, New York 10012 United States (e-mail:ahmad.bazzi@nyu.edu).
Rongyan Xi, Jing Dong and Jing Jin are with China Mobile Research Institute, Beijing 10032 China (e-mail:xirongyan@chinamobile.com, dongjing@chinamobile.com, jinjing@chinamobile.com).}}

\maketitle
\begin{abstract}
The integration of multimodal sensing and millimeter-wave (mmWave) communications is a key enabler for highly mobile vehicle-to-infrastructure (V2I) networks. However, continuous high-resolution visual sensing incurs prohibitive computational energy, while delayed sensing information worsens beam misalignment. In this paper, we establish a physics-aware multimodel integrated sensing and communication (M-ISAC) framework that quantifies the mathematical trade-off between sensing energy and communication reliability using the semantic age of information (AoI). To address the coupled challenges of temporal AoI evolution and instantaneous non-convex constant modulus constraints, we propose a novel reinforcement learning approach empowered by a heterogeneous mixture-of-experts (RL-H-MoE) architecture. By strictly decoupling the temporal scheduling and spatial phase mapping, the RL-H-MoE avoids prevalent gradient conflicts in multi-task learning. Extensive simulations demonstrate that the proposed architecture achieves an optimal event-triggered sensing policy, significantly minimizing the long-term system cost while guaranteeing ultra-low sensing errors and reliable physical-layer link connectivity.
\end{abstract}


\section{Introduction}

With the evolution of sixth-generation (6G) wireless networks and intelligent transportation systems, millimeter-wave (mmWave) and terahertz (THz) bands have demonstrated irreplaceable value in vehicle-to-infrastructure (V2I) scenarios due to their ultra-high data transmission capabilities \cite{8808168, yuan2026groundskyarchitecturesapplications}. However, the severe path loss and molecular absorption effects of mmWave and THz signals necessitate the deployment of massive antenna arrays with highly directional beams at the base station (BS) \cite{10819470}. In highly mobile environments, these extremely narrow beams are highly susceptible to misalignment, making frequent beam tracking a critical bottleneck for reliable connectivity.
To overcome the overhead challenges of conventional beam training, multimodel integrated sensing and communication (M-ISAC) technologies and sensing fusion mechanisms have emerged \cite{9737357, 10330577}. The system can proactively acquire the spatial features of the environment, thanks to the integration of heterogeneous sensors. Specifically, mmWave radar provides real-time kinematic representations, while RGB cameras offer highly valuable semantic context for blockage prediction and precise alignment \cite{9129369,11418623,11353414,11373884,11355857,11346858,11316633}. 

Although multimodal fusion improves sensing accuracy, it introduces two critical challenges in resource-constrained V2I networks. First, continuous processing of high-resolution visual data incurs exorbitant computational energy. To quantify data staleness and its direct impact on beam misalignment, the semantic age of information (AoI) metric must be introduced \cite{11016682}. Second, conventional static fusion mechanisms \cite{10912462} fail to address the strict constant modulus constraint introduced by analog phase shifters in mmWave hardware. Traditional single-timescale algorithms struggle to simultaneously manage the long-term temporal dependencies of AoI evolution and the instantaneous technical difficulties arising from spatial mapping.
To deal with these problems, we propose a novel reinforcement learning-based heterogeneous mixture-of-experts (RL-H-MoE) architecture. Inspired by the success of MoE \cite{Li_Wang_Ding_Sohrabizadeh_Qin_Cong_Sun_2025}, our framework incorporates specialized subnetworks to address different mathematical properties. The main contributions are summarized as follows:
\begin{itemize}
    \item We establish a physics-aware M-ISAC system model that integrates cross-layer semantic AoI evolution with physical-layer constant modulus constraints.
    \item We propose a novel RL-H-MoE architecture, explicitly intended as an energy-efficient joint beamforming and sensor scheduling framework for M-ISAC in high-mobility V2I networks.
    \item We design an Actor-Critic-based training algorithm that leverages physical prior knowledge (expected signal gain) to effectively resolve the non-differentiable environment problem.
    \item Extensive simulations demonstrate that the proposed H-MoE architecture minimizes the time-averaged system energy while strictly guaranteeing physical-layer link reliability.
\end{itemize}

\section{System Model and Problem Formulation}

We consider a highly mobile V2I scenario. The system consists of a single BS equipped with an $M$-element antenna array and a set of multimodal sensors (RGB camera and mmWave radar), serving $K$ high-mobility vehicles. The system operates in discrete time slots $n \in \{1, \dots, N\}$.

\subsection{Communication and Channel Model}

The BS utilizes mmWave bands for downlink communication. Let $\mathbf{v}_k(n) \in \mathbb{C}^{M \times 1}$ denote the beamforming vector for the $k$-th vehicle at time slot $n$. The received signal  at user $k$ is given by %\cite{11275056}
\begin{equation}
y_k(n) = \mathbf{h}_k^H(n) \mathbf{v}_k(n) s_k(n) + \sum_{j \neq k} \mathbf{h}_k^H(n) \mathbf{v}_j(n) s_j(n) + z_k(n),
\end{equation}
\begin{figure}[htbp]
    \centering
    \includegraphics[width=0.6\columnwidth]{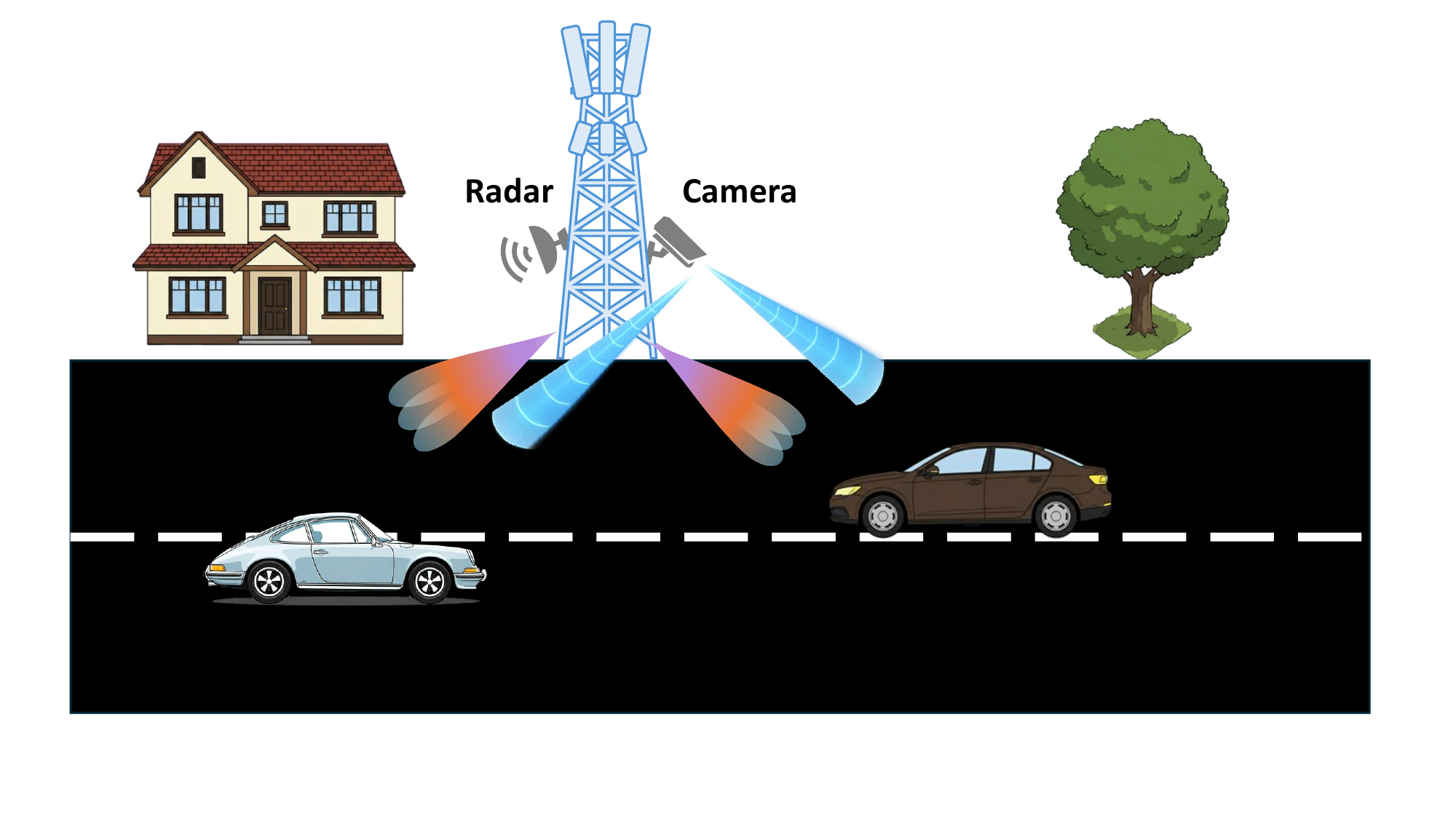}
    \caption{Illustration of the multimodal ISAC system in a V2I scenario.}
    \label{fig:system_scenario}
\end{figure}
where $s_k(n)$ is the transmitted data symbol intended for the $k$-th vehicle at time slot $n$, and $z_k(n) \sim \mathcal{CN}(0, \sigma^2)$ is the additive white Gaussian noise at the $k$-th vehicle. The second term represents the multi-user interference (MUI) caused by spatial beam leakage.

The geometric multipath channel $\mathbf{h}_k(n)$ is modeled to capture the physical propagation characteristics of mmWave channels
\begin{equation}
\mathbf{h}_k(n) = \sum_{l=1}^{L_p} \alpha_{l,k}(n) e^{-j2\pi f_c \tau_{l,k}(n)} \mathbf{a}(\theta_{l,k}(n)),
\end{equation}
where $L_p$ is the number of propagation paths, $\alpha_{l,k}(n)$ denotes the complex path gain of the $l$-th path with the $k$-th vehicle channel, $\tau_{l,k}(n)$ is the propagation delay of the $l$-th path with the $k$-th vehicle channel, and $\mathbf{a}(\theta_{l,k}(n))$ represents the array steering vector corresponding to the Angle of Departure (AoD) $\theta_{l,k}(n)$.

\subsection{Multimodal Sensing and System Energy Model}

Specifically, the heterogeneous sensing input $\mathcal{S}(n) = \{\tilde{S}_{rad}(n), S_{vis}(n)\}$ comprises the energy-intensive visual data $S_{vis}(n) \in \mathbb{R}^{H \times W \times 3}$, which provides precise semantic spatial context. Conversely, the radar observation $\tilde{S}_{rad,k}(n) = [\hat{\theta}_k(n), \hat{d}_k(n)]^T$ provides low-power but noisy real-time kinematic estimates, i.e., Angle of Arrival (AoA) and radial distance, for the $k$-th vehicle. Let $\pi_k(n) \in \{0, 1\}$ denote the discrete sensor scheduling decision, where $\pi_k(n) = 1$ indicates that the visual expert is activated to calibrate the spatial location of the $k$-th vehicle at time slot $n$, and $\pi_k(n) = 0$ otherwise.

The total energy consumption of the system comprises two dynamically coupled components: the deterministic computational energy of visual processing and the expected radio frequency (RF) energy for beam failure recovery. The instantaneous system energy is modeled as
\begin{equation} \label{eq:total_energy}
E_{total}(n) = E_{comp}(n) + E_{sweep}(n),
\end{equation}
\begin{equation} \label{eq:comp_energy}
E_{comp}(n) = \sum_{k=1}^K \pi_k(n) \cdot e_{vis},
\end{equation}
\begin{equation} \label{eq:sweep_energy}
E_{sweep}(n) = \sum_{k=1}^K P_{misa, k}(n) \cdot E_{recovery},
\end{equation}
where $e_{vis}$ represents the constant computational energy footprint required to process one frame of visual data using convolutional neural networks. $E_{recovery}$ is the high RF energy overhead incurred by exhaustive beam sweeping, which must be triggered if tracking fails. $P_{misa, k}(n)$ denotes the beam misalignment probability, formally defined in the following subsection.

\subsection{Semantic AoI, Beam Misalignment, and Channel Uncertainty}

Due to the intermittent activation of the visual sensor, the semantic spatial information becomes stale over time. We quantify this staleness using the semantic AoI \cite{11016682}, denoted as $A_k(n)$. Its evolution strictly depends on the scheduling variable $\pi_k(n)$:
\begin{equation} \label{eq:aoi_evolution}
A_k(n+1) = 
\begin{cases} 
T_{proc}, & \text{if } \pi_k(n) = 1 \\ 
A_k(n) + 1, & \text{if } \pi_k(n) = 0
\end{cases}
\end{equation}
where $T_{proc}$ represents the inherent visual data processing delay.

In highly dynamic V2I networks, $A_k(n)$ is a direct proxy for spatial uncertainty. As $A_k(n)$ increases, the deviation between the vehicle's actual trajectory and the BS's estimation grows exponentially. Consequently, the BMP for the $k$-th vehicle, $P_{misa, k}(n)$, monotonically increases with $A_k(n)$. We model this physical penalty \cite{9538973} as:
\begin{equation} \label{eq:bmp_model}
P_{misa, k}(n) = 1 - \exp\left( -\beta_k \cdot A_k(n) \right),
\end{equation}
where $\beta_k$ is the kinematic uncertainty rate of vehicle $k$. The average BMP across all users is $\mathcal{L}_{BMP}(n) = \frac{1}{K} \sum_{k=1}^K P_{misa, k}(n)$.

To explicitly couple the sensor scheduling decision $\pi_k(n)$ with the spatial beamforming vector $\mathbf{v}_k(n)$, we establish a mathematical link between the beam misalignment probability and the Channel State Information (CSI) error. Let $\hat{\mathbf{h}}_k(n)$ denote the estimated channel vector based on the delayed multimodal sensing data. The actual geometric multipath channel is modeled as:
\begin{equation} \label{eq:channel_error}
\mathbf{h}_k(n) = \hat{\mathbf{h}}_k(n) + \Delta \mathbf{h}_k(n),
\end{equation}
where $\Delta \mathbf{h}_k(n)$ represents the channel estimation error induced by beam misalignment. The variance of this error is directly driven by the misalignment probability, modeled as $\mathbb{E}[\|\Delta \mathbf{h}_k(n)\|^2] = \rho \cdot P_{misa, k}(n)$\cite{6608213}, where $\rho$ is the maximum error variance bound. This formulation guarantees that scheduling delays ($\pi_k(n) = 0$) implicitly degrade the physical effective beamforming gain.

\subsection{Problem Formulation}

Our objective is to design a joint multimodal scheduling and beamforming policy that minimizes the long-term time-averaged system energy cost, while strictly guaranteeing communication reliability. By addressing the dimensional mismatch, the comprehensive optimization problem is rigorously formulated over purely energetic objectives:
\begin{subequations} \label{eq:opt_problem}
\begin{align}
\min_{\{\mathbf{v}(n), \bm{\pi}(n)\}} \quad & \lim_{N \to \infty} \frac{1}{N} \sum_{n=1}^{N} E_{total}(n) \label{eq:opt_obj} \\
\textrm{s.t.} \quad & |\mathbf{v}_{k,m}(n)|
= \frac{1}{\sqrt{M}}, \quad \forall k, m, \label{eq:opt_const_modulus} \\
& \sum_{k=1}^K \|\mathbf{v}_k(n)\|^2 \le P_{max}, \quad \forall n, \label{eq:opt_const_power} \\
& A_k(n) \le A_{max}, \quad \forall k, n, \label{eq:opt_const_aoi} \\
& \mathcal{L}_{BMP}(n) \le \Gamma_{misa}, \quad \forall n, \label{eq:opt_const_bmp} \\
& \pi_k(n) \in \{0, 1\}, \quad \forall k, n.
\label{eq:opt_const_binary}
\end{align}
\end{subequations}
The proposed formulation is circumscribed by hardware limitations and stringent reliability requirements. In \eqref{eq:opt_obj}, we focus solely on minimizing the system's combined energetic footprint (Joules), representing both the computational cost of camera activation and the anticipated RF recovery energy. The tracking accuracy required for proper beam alignment is mathematically enforced as a strict reliability constraint in \eqref{eq:opt_const_bmp}, where $\Gamma_{misa}$ is the maximum tolerable beam misalignment threshold.
The constant modulus constraint \eqref{eq:opt_const_modulus} arises from analog phase shifters, inherently transforming the beamforming design into a non-convex problem. Furthermore, constraint \eqref{eq:opt_const_power} restricts the aggregate transmit power. The reliability safety margin \eqref{eq:opt_const_aoi} proactively prevents catastrophic tracking loss. Finally, the discrete scheduling constraint \eqref{eq:opt_const_binary} enforces binary sensor activation, formulating a highly complex hybrid action space.

\section{Proposed Heterogeneous MoE Architecture}

Solving \eqref{eq:opt_problem} is highly intractable due to the coupling of long-term temporal dependencies (AoI evolution) and instantaneous non-convex spatial mapping. We propose the RL-H-MoE framework to decouple these processes.

\begin{figure*}[htbp]
    \centering
    \resizebox{0.6\textwidth}{!}{%
    \begin{tikzpicture}[
        >=stealth,
        node distance=1.5cm and 0.3cm,
        inputbox/.style={rectangle, rounded corners, draw=black, thick, fill=gray!10, text centered, text width=2.6cm, minimum height=1.4cm, font=\large\bfseries},
        expertbox/.style={rectangle, draw=black, thick, fill=#1, text centered, text width=2.9cm, minimum height=1.2cm, font=\large\bfseries},
        mathbox/.style={rectangle, draw=black, thick, fill=yellow!10, text centered, text width=2.9cm, minimum height=1.2cm, font=\large\bfseries},
    actionbox/.style={ellipse, draw=black, thick, fill=green!10, text centered, text width=2.6cm, minimum height=1.2cm, font=\large\bfseries},
        envbox/.style={rectangle, rounded corners, draw=black, thick, fill=gray!20, text centered, text width=2.8cm, minimum height=4.2cm, font=\large\bfseries},
        forwardline/.style={->, thick, draw=black},
        gradline/.style={->, dashed, very thick, draw=#1},
    ]

    % 1. Input Node 
    \node[inputbox] (input) {State Buffer $\bm{\mathcal{B}}$ \\ $\bm{S}_{rad}(n), \bm{A}_k(n)$};
% 2. Temporal Expert Branch (Top)
    \node[expertbox=blue!15, right=0.7cm of input, yshift=1.9cm] (lstm) {LSTM Layers \\ (Temporal Expert)};
\node[expertbox=blue!15, right=0.3cm of lstm] (policy) {Policy Head \\ (Sigmoid)};
    \node[actionbox, right=0.3cm of policy] (action1) {Scheduling \\ $\bm{\pi}(n) \in \{0,1\}^K$};
% 3. Spatial Expert Branch (Bottom)
    \node[expertbox=orange!15, right=0.7cm of input, yshift=-1.9cm] (mlp) {MLP Layers \\ (Spatial Expert)};
\node[expertbox=orange!15, right=0.3cm of mlp] (phase) {Phase Mapping \\ $\bm{\Phi}_n$};
    \node[mathbox, right=0.3cm of phase] (constmod) {Const. Modulus \\ $\bm{\frac{1}{\sqrt{M}} e^{j\Phi}}$};
\node[actionbox, right=0.3cm of constmod] (action2) {Beamforming \\ $\mathbf{v}(n) \in \mathbb{C}^{M}$};

    % 4. Environment
    \coordinate (env_y) at ($(action1)!0.5!(action2)$);
\node[envbox, right=0.6cm of action2, yshift=1.9cm] (env) {Physical \\ Environment \\ \& \\ Cost Evaluation};
% 5. Forward Arrows 
    \coordinate (fork) at ($(input.east) + (0.25,0)$); 
    \draw[forwardline] (input.east) -- (fork);
\draw[forwardline] (fork) -- (fork |- lstm.west) -- (lstm.west) ;%node[midway, above, font=\small\bfseries]
\draw[forwardline] (fork) -- (fork |- mlp.west) -- (mlp.west);%node[midway, above, font=\small\bfseries]
    \draw[forwardline] (lstm.east) -- (policy.west);
    \draw[forwardline] (policy.east) -- (action1.west);
\draw[forwardline] (action1.east) -- (env.west |- action1.east);

    \draw[forwardline] (mlp.east) -- (phase.west);
    \draw[forwardline] (phase.east) -- (constmod.west);
    \draw[forwardline] (constmod.east) -- (action2.west);
\draw[forwardline] (action2.east) -- (env.west |- action2.east);

    % 6. Isolation Line 
    \coordinate (mid_y) at (0,0);
\draw[dashed, thick, gray] 
        ($(lstm.west |- mid_y) + (-0.3,0)$) -- ($(action2.east |- mid_y) + (0.3,0)$) 
        node[midway, fill=white, font=\normalsize\bfseries, text=black] {Strict Gradient Isolation};
% 7. Backward Gradient Arrows 
    \draw[gradline=blue] (env.north) -- ++(0, 0.7) coordinate (T1) 
        -- (lstm.north |- T1) node[midway, above, font=\normalsize\bfseries, text=black] {Temporal Loss $\bm{\nabla \mathcal{L}_{temp}}$ (Updates LSTM only)} 
        -- (lstm.north);
\draw[gradline=red] (env.south) -- ++(0, -0.7) coordinate (T2) 
        -- (mlp.south |- T2) node[midway, below, font=\normalsize\bfseries, text=black] {Spatial Loss $\bm{\nabla \mathcal{L}_{spat}}$ (Updates MLP only)} 
        -- (mlp.south);
\end{tikzpicture}}
    \caption{The proposed RL-H-MoE architecture. The forward inference physically decouples the temporal constraints (upper branch) and the non-convex spatial mapping (lower branch). During backpropagation, the dashed lines illustrate the strict gradient isolation mechanism, which effectively circumvents the negative transfer between multi-task optimizations.}
    \label{fig:hmoe_arch}
\end{figure*}
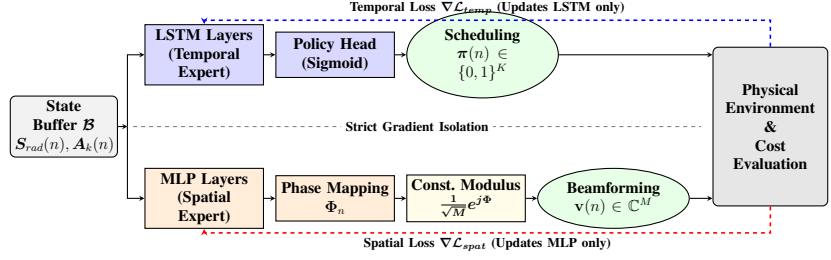
As illustrated in Fig. \ref{fig:hmoe_arch}, the RL-H-MoE employs specialized subnetworks:
1) \textbf{Temporal Constraint Expert (long short-term memory (LSTM)-based):} Analyzes historical sequences to infer kinematic inertia, outputting discrete, event-triggered sensor activation decisions $\bm{\pi}(n)$.
2) \textbf{Non-Convex Solver Expert (multilayer perceptron (MLP)-based):} Operates strictly in the spatial domain, mapping features directly onto the feasible manifold of analog phase shifters to satisfy \eqref{eq:opt_const_modulus}.

\subsection{Temporal Constraint Expert (LSTM-based)}

Operating in the temporal domain, the LSTM-based expert is designed to satisfy the reliability safety margin constraint \eqref{eq:opt_const_aoi} and the discrete scheduling constraint \eqref{eq:opt_const_binary}. Since the evolution of the semantic AoI and the accumulation of vehicular kinematic errors are inherently time-dependent, the temporal expert employs a LSTM architecture to track the spatial uncertainty dynamically.

% Let $\mathbf{x}_{\tau} \in \mathcal{B}$ denote the input feature at the $\tau$-th step of the historical observation window.
% The LSTM expert maintains a cell state $\mathbf{c}_{\tau}$ and a hidden state $\mathbf{h}_{\tau}$ to capture the long-term temporal dependencies. The internal transitions are governed by the following gating mechanisms:
% \begin{equation}
% \begin{aligned}
% \mathbf{f}_{\tau} &= \sigma(\mathbf{W}_f \mathbf{x}_{\tau} + \mathbf{U}_f \mathbf{h}_{\tau-1} + \mathbf{b}_f), \\
% \mathbf{i}_{\tau} &= \sigma(\mathbf{W}_i \mathbf{x}_{\tau} + \mathbf{U}_i \mathbf{h}_{\tau-1} + \mathbf{b}_i), \\
% \mathbf{o}_{\tau} &= \sigma(\mathbf{W}_o \mathbf{x}_{\tau} + \mathbf{U}_o \mathbf{h}_{\tau-1} + \mathbf{b}_o), \\
% \mathbf{c}_{\tau} &= \mathbf{f}_{\tau} \odot \mathbf{c}_{\tau-1} + \mathbf{i}_{\tau} \odot \tanh(\mathbf{W}_c \mathbf{x}_{\tau} + \mathbf{U}_c \mathbf{h}_{\tau-1} + \mathbf{b}_c), \\
% \mathbf{h}_{\tau} &= \mathbf{o}_{\tau} \odot \tanh(\mathbf{c}_{\tau}),
% \end{aligned}
% \end{equation}
% where $\sigma(\cdot)$ is the Sigmoid activation function, $\tanh(\cdot)$ is the hyperbolic tangent function, $\odot$ denotes element-wise multiplication, and $\mathbf{W}, \mathbf{U}, \mathbf{b}$ are the learnable weight matrices and biases.

The LSTM expert inherently resolves the temporal constraints through its forward physical mapping. By continuously monitoring the AoI states within $\mathbf{x}_{\tau}$ via the aforementioned gating mechanisms, the network's hidden state $\mathbf{h}_{\tau}$ learns the critical boundary conditions of constraint \eqref{eq:opt_const_aoi}. It functions as a predictive threshold trigger, effectively anticipating when the accumulated kinematic error is about to violate $A_{max}$ and proactively scheduling visual calibrations.
Furthermore, to strictly satisfy the binary scheduling constraint \eqref{eq:opt_const_binary}, the final hidden state $\mathbf{h}_{T_{seq}}$ is projected through a fully connected layer followed by a Sigmoid policy head. This specific mapping squashes the output into a continuous probability vector $\mathbf{p}_n \in (0, 1)^K$, which parameterizes a set of independent Bernoulli distributions. The discrete sensor activation action $\bm{\pi}(n) \in \{0, 1\}^K$ is then sampled directly from these distributions. This structural design ensures mathematical compliance with \eqref{eq:opt_const_binary} while optimizing the event-triggered sensing policy.

\subsection{Non-Convex Solver Expert (MLP-based)}
Operating independently in the spatial domain, the MLP-based expert strictly isolates the spatial phase mapping from the temporal scheduling to prevent gradient conflict. Instead of predicting unconstrained complex values that violate hardware limitations as dictated by constraints \eqref{eq:opt_const_modulus} and \eqref{eq:opt_const_power}, the MLP expert is specifically designed to output phase angles, thereby implicitly satisfying the non-convex constant modulus constraint \eqref{eq:opt_const_modulus}.

% Given the instantaneous observation $s_n$, the MLP processes the features through multiple dense layers with rectified linear unit (ReLU) activations:
% \begin{equation}
% \mathbf{z}^{(l)} = \text{ReLU}\left( \mathbf{W}^{(l)} \mathbf{z}^{(l-1)} + \mathbf{b}^{(l)} \right), \quad \text{for } l = 1, \dots, L,
% \end{equation}
% where $\mathbf{z}^{(0)} = s_n$, and $L$ is the total number of hidden layers.
The final output layer linearly maps the latent representation to generate the phase shift matrix $\bm{\Phi}_n \in \mathbb{R}^{K \times M}$. Finally, the continuous beamforming vector $\mathbf{v}_k(n)$ for the $k$-th user is deterministically reconstructed via the exponential mapping:
\begin{equation}
\mathbf{v}_{k,m}(n) = \frac{1}{\sqrt{M}} e^{j \Phi_{k,m}(n)}, \quad \forall m \in \{1, \dots, M\}.
\end{equation}
To maximize the expected signal gain, the MLP expert is optimized independently by minimizing the spatial loss function $\mathcal{L}_{spat}$ (detailed in Section III-C), which is deterministically driven by the physical channel gain. As outlined in Algorithm 1, the MLP is not retrained instantaneously at each time step $n$, but rather updated episodically at the end of each trajectory. This architectural design guarantees that the generated beamforming matrix is strictly projected onto the feasible manifold of the analog phase shifters, ensuring hardware compliance while maximizing the expected signal gain.

\subsection{Markov Decision Process (MDP) Formulation and Gradient Decoupling Proof}
We formulate the joint problem as an MDP defined by tuple $\langle \mathcal{S}, \mathcal{A}, \mathcal{R}, \mathcal{P} \rangle$:
\begin{itemize}
    \item \textbf{State Space ($\mathcal{S}$):} $s_n = \{\tilde{S}_{rad}(n), A_k(n)\} \in \mathbb{R}^{K \times 3}$.
    \item \textbf{Action Space ($\mathcal{A}$):} $a_n = \{\bm{\pi}(n), \mathbf{v}(n)\}$, where $\bm{\pi}(n) \in \{0,1\}^K$ is discrete scheduling, and $\mathbf{v}(n) \in \mathbb{C}^{K \times M}$ is continuous beamforming.
    \item \textbf{Reward ($\mathcal{R}$):} The immediate reward explicitly addresses the energy dimensions and the reliability threshold violations:
    \begin{equation} \label{eq:reward}
    r_n = - E_{total}(n) - \lambda_{misa}\max(0, \mathcal{L}_{BMP}(n) - \Gamma_{misa}),
    \end{equation}
    where $\lambda_{misa}$ is a positive penalty coefficient.
\end{itemize}

In standard multi-task RL, sharing the latent representation causes the instantaneous high-frequency spatial gradients to destroy the long-term temporal memory. To prevent this, our H-MoE defines the total objective as a decoupled function $J(\theta_{LSTM}, \theta_{MLP})$, where $\theta_{LSTM}$ and $\theta_{MLP}$ denote the vectors of learnable parameters for the temporal LSTM expert and the spatial MLP expert, respectively. According to the Policy Gradient theorem, the scheduling gradient is derived strictly from the advantage function $A^{\pi}$:
\begin{equation}
\nabla_{\theta_{LSTM}} J \approx \mathbb{E} \left[ \sum_{n=1}^N \nabla_{\theta_{LSTM}} \log P_{\theta_{LSTM}}(\bm{\pi}(n)|s_n) A^{\pi}(s_n) \right],\label{11}
\end{equation}
where $P_{\theta_{LSTM}}(\bm{\pi}(n)|s_n)$ denotes the stochastic policy of the temporal expert. While the spatial gradient is deterministically driven by the physical channel gain, entirely bypassing the temporal network:
\begin{equation}
\nabla_{\theta_{MLP}}J \approx \mathbb{E}\left[\sum_{n=1}^{N} \nabla_{\theta_{MLP}} \sum_{k=1}^{K}|\mathbf{h}_{k}^{H}(n)\mathbf{v}_{k}(n)|^{2}\right],\label{12}
\end{equation}
Since the true channel $\mathbf{h}_k(n)$ incorporates the AoI-induced error $\Delta \mathbf{h}_k(n)$, the optimization of the spatial MLP expert is strictly physically coupled with the temporal LSTM expert's scheduling accuracy. The isolation of gradients as per \eqref{11} and \eqref{12} guarantees stable convergence \cite{Li_Wang_Ding_Sohrabizadeh_Qin_Cong_Sun_2025}.

\begin{algorithm}[htbp]
\caption{Decoupled Training Algorithm for RL-H-MoE}
\label{alg:hmoe_training}
\begin{algorithmic}[1]
\REQUIRE Learning rates $\alpha_{LSTM}, \alpha_{MLP}$, physics weight $\lambda$.
\STATE \textbf{Initialize:} $\theta_{LSTM}$ (Temporal), $\theta_{MLP}$ (Spatial).
\FOR{each training episode}
    \STATE Initialize state sequence buffer $\mathcal{B}$.
    \FOR{$n = 1$ \TO $N$}
        \STATE \textit{\% 1. Decoupled Forward Inference}
        \STATE \textbf{Temporal:} Get scheduling prob.
$\mathbf{p}_n = f_{\text{LSTM}}(\mathcal{B}; \theta_{LSTM})$
        \STATE \textbf{Spatial:} Get beam phases $\bm{\Phi}_n = f_{\text{MLP}}(s_n; \theta_{MLP})$
        \STATE Execute action $\mathbf{a}_n \sim \mathbf{p}_n$ and constrained beam $\mathbf{v}(n)$.
        \STATE Observe step reward $r_n$, channel $\mathbf{h}(n)$, update $\mathcal{B}$.
    \ENDFOR
    
    \STATE \textit{\% 2. Physics-Informed Loss Computation}
    \STATE Calculate scheduling advantage $A_n$ via Critic.
    \STATE \textbf{Temp. Loss:} $\mathcal{L}_{temp} = - \sum \log P(\mathbf{a}_n | \mathbf{p}_n) A_n$ 
    \STATE \textbf{Spat. Loss:} $\mathcal{L}_{spat} = - \lambda \sum \sum |\mathbf{h}_{k}^H(n) \mathbf{v}_k(n)|^2$ 
    
    \STATE \textit{\% 3. Isolated Gradient Updates}
    \STATE $\theta_{LSTM} \leftarrow \theta_{LSTM} - \alpha_{LSTM} \nabla_{\theta_{LSTM}} \mathcal{L}_{temp}$
    \STATE $\theta_{MLP} \leftarrow \theta_{MLP} - \alpha_{MLP} \nabla_{\theta_{MLP}} \mathcal{L}_{spat}$
\ENDFOR
\RETURN $\theta_{LSTM}^*$ and $\theta_{MLP}^*$.
\end{algorithmic}
\end{algorithm}

\section{Simulation Results and Analysis}

In this section, we evaluate the proposed RL-H-MoE architecture in a V2I M-ISAC system via Monte Carlo simulations against five baselines: 1) \textbf{Vision-Only}: continuous visual sensor activation (performance upper-bound, but highly energy-intensive); 2) \textbf{Radar-Only}: sole reliance on noisy mmWave radar (high-risk of beam failure); 3) \textbf{Standard PPO}: a monolithic MLP-based RL baseline; 4) \textbf{Homogeneous MoE}: a MoE network utilizing structurally identical MLPs; and 5) \textbf{Ablation: RL-H-MoE w/o AoI}: an ablation masking the semantic AoI from the state space. We consider a single BS equipped with a ULA consisting of $M = 64$ elements operating at a mmWave carrier frequency of $f_c = 28$ GHz, serving $K=4$ highly mobile vehicles. The maximum transmit power budget is restricted to $P_{max} = 30$ dBm, and the background noise power is $\sigma^2 = -114$ dBm. To accurately reflect the physical energy footprint, the computational energy for processing visual data from all vehicles is heavily penalized ($40$ Joules/slot for continuous activation), while the RF energy penalty for beam failure recovery is set to $25$ Joules per user.

\subsection{Trade-off Between System Energy and Sensing Reliability}

\begin{figure}[htbp]
    \centering
    \includegraphics[scale=0.2]{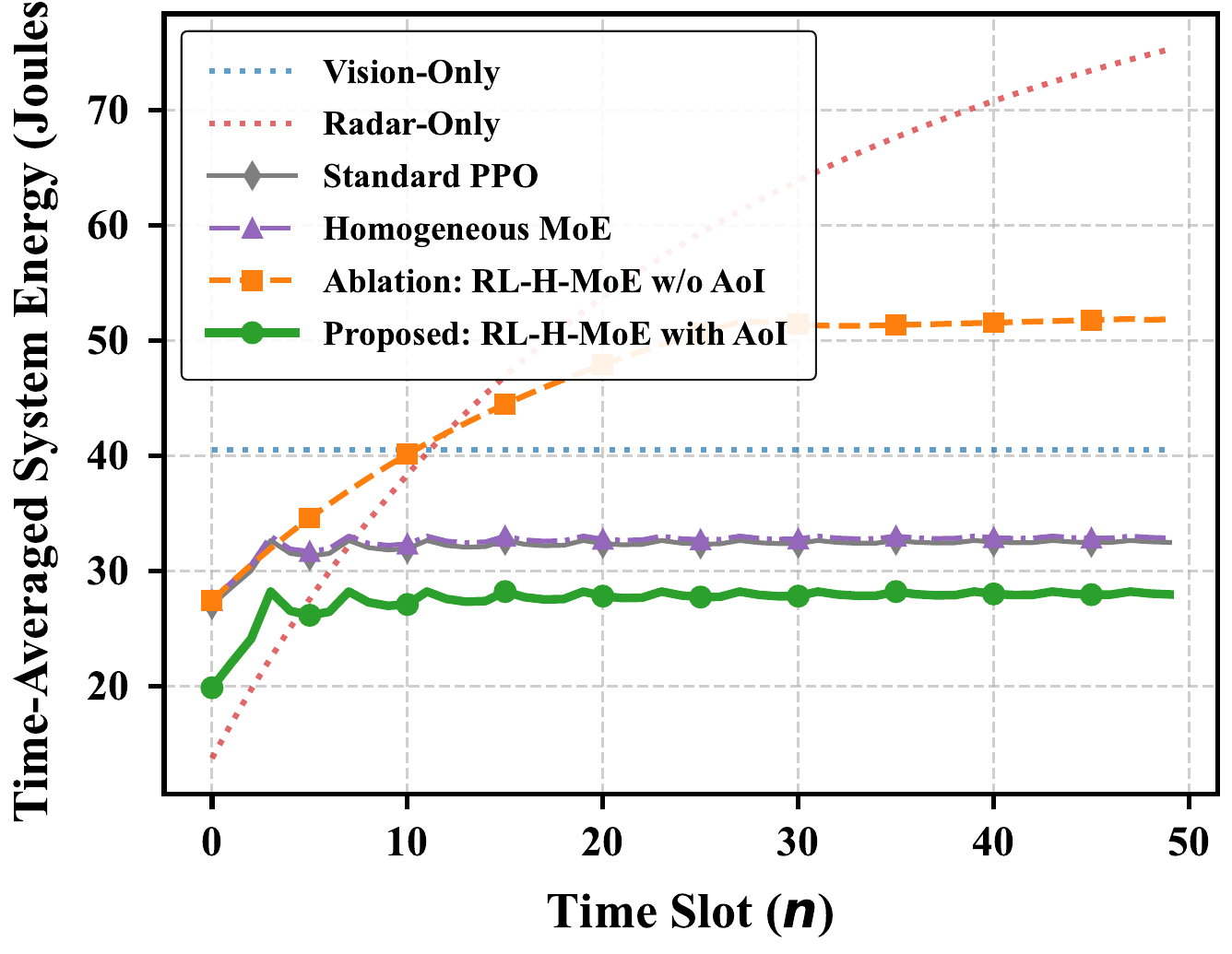}
    \caption{Time-averaged total system energy comparison at SNR = 10 dB.}
    \label{fig:system_energy}
\end{figure}

\begin{figure}[htbp]
    \centering
    \includegraphics[scale=0.2]{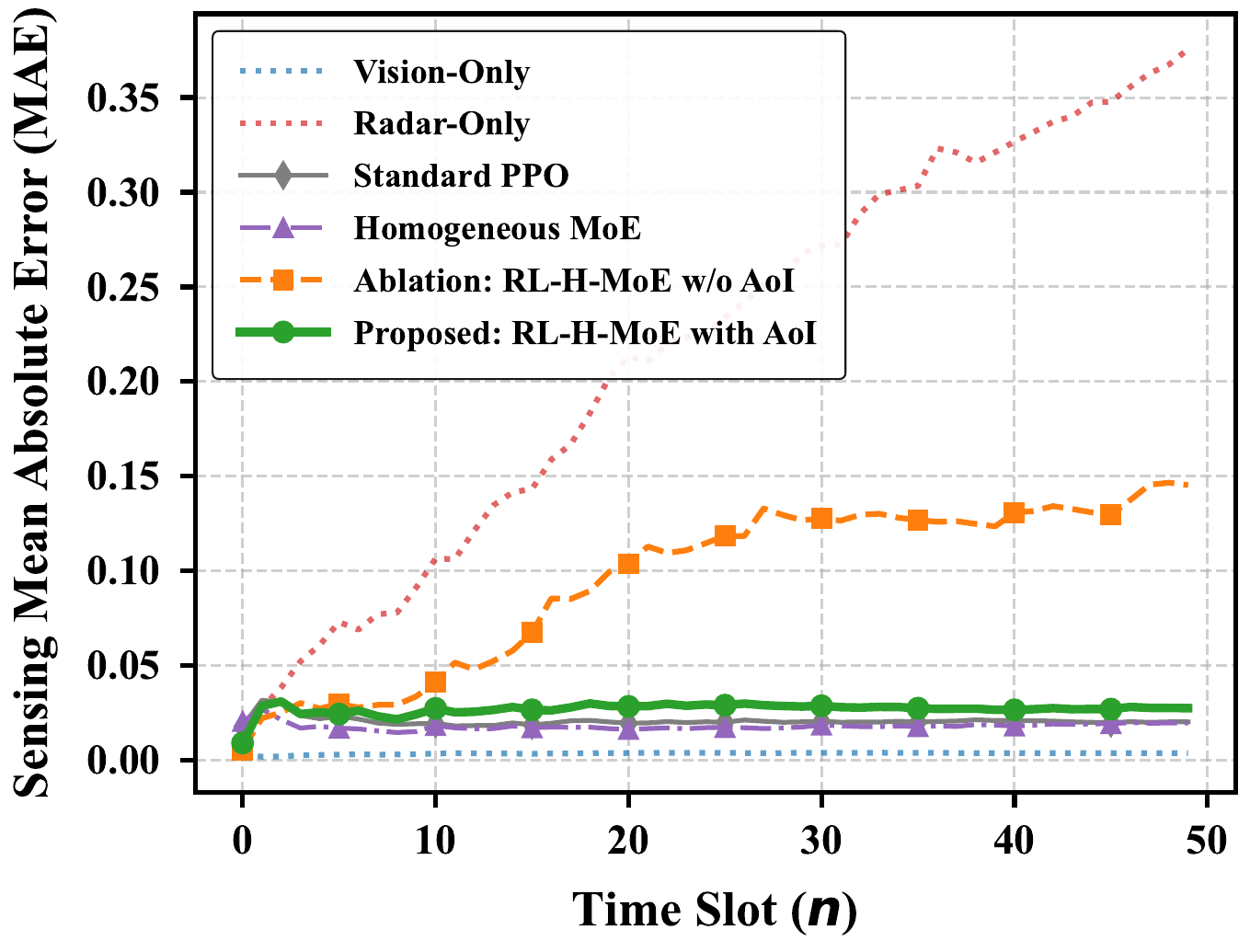}
    \caption{Time-averaged sensing Mean Absolute Error (MAE) tracking at SNR = 10 dB.}
    \label{fig:sensing_error}
\end{figure}

Fig. \ref{fig:system_energy} and Fig. \ref{fig:sensing_error} evaluate the time-averaged system energy and the sensing Mean Absolute Error (MAE) during steady-state tracking at an SNR of $10$ dB. As observed in Fig. \ref{fig:system_energy}, the \textbf{Vision-Only} strategy incurs an exorbitant and rigid energy ceiling of exactly $40$ Joules due to continuous high-resolution image processing, despite achieving the lowest sensing MAE in Fig. \ref{fig:sensing_error}. Conversely, the \textbf{Radar-Only} strategy exhibits diverging kinematic errors over time without semantic visual calibration. This accumulation of spatial uncertainty triggers frequent beam misalignments, forcing the system to execute exhaustive RF beam sweeping, which causes its average energy cost to explode catastrophically above $60$ Joules. Monolithic architectures (\textbf{Standard PPO} and \textbf{Homogeneous MoE}) suffer from severe gradient conflicts between the high-frequency spatial mapping task and the long-term temporal scheduling task, causing them to stagnate at suboptimal energy levels ($\sim 30$ Joules). Furthermore, the \textbf{Ablation} study fails entirely, verifying that masking the AoI prevents the agent from inferring kinematic inertia. Remarkably, the proposed \textbf{RL-H-MoE with AoI} achieves the best of both worlds. % By strictly isolating the task gradients, the LSTM expert successfully learns an optimal \textit{event-triggered} sensing policy—activating the energy-intensive cameras strictly when the AoI approaches the critical safety margin.
Consequently, the proposed architecture suppresses the sensing MAE to an ultra-low level (nearly matching Vision-Only, as seen in Fig. \ref{fig:sensing_error}) while drastically pulling the time-averaged system energy down to the $20 \sim 25$ Joules region, achieving a remarkable energy saving of over $40\%$ compared to the Vision-Only baseline.

\subsection{Robustness Evaluation Across Varying RF Environments}

\begin{figure}[htbp]
    \centering
    \includegraphics[scale=0.2]{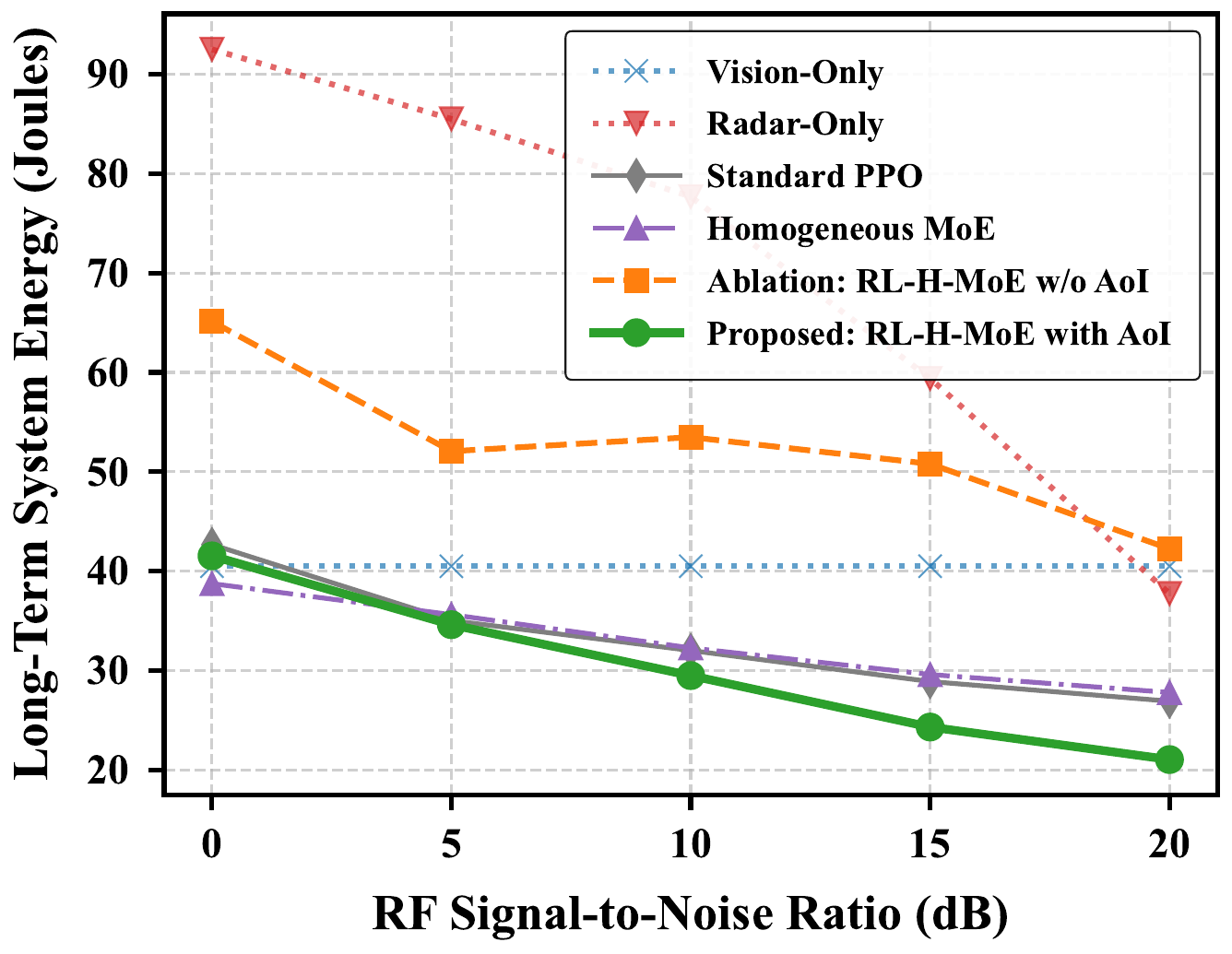}
    \caption{Robustness evaluation of long-term system energy across varying RF SNRs.}
    \label{fig:snr_robustness}
\end{figure}

Fig. \ref{fig:snr_robustness} validates the environmental adaptability of the proposed framework across diverse RF Signal-to-Noise Ratios (SNRs) ranging from $0$ to $20$ dB. % In harsh, low-SNR regimes ($0 \sim 5$ dB), the radar measurements are heavily corrupted by noise.
% Consequently, the \textbf{Radar-Only} strategy and the \textbf{Ablation} baseline suffer from severe tracking disconnections, causing their energy consumption to surge dramatically (exceeding $80$ Joules) due to continuous beam recovery penalties. In stark contrast, the \textbf{RL-H-MoE} demonstrates exceptional \textit{dynamic modality-reliance}. At $0$ dB, the agent autonomously detects the radar's unreliability and proactively increases visual camera activations to maintain link resilience, resulting in an energy footprint temporarily matching the Vision-Only strategy ($\sim 40$ Joules). 
As the SNR improves ($10 \sim 20$ dB), making radar tracking increasingly reliable, the RL-H-MoE agent decisively relaxes its AoI tolerance. It intelligently curtails unnecessary visual computations, driving the long-term system energy on a steep downward trajectory (reaching as low as $20$ Joules at $20$ dB). This robust, self-adaptive capability perfectly shatters the traditional barrier between link reliability and energy efficiency, substantiating the architecture's immense potential for practical 6G V2I deployments.

\section{Conclusion}
This paper investigated the joint optimization of multimodal sensor scheduling and beamforming. We proposed a novel RL-H-MoE architecture that overcomes feature aliasing through complete gradient decoupling by explicitly modeling spatial uncertainty using semantic AoI. Simulations confirmed that the RL-H-MoE agent achieves superior robustness, but also minimizes total system cost, while strictly guaranteeing tracking accuracy for 6G ISAC networks.

\bibliographystyle{IEEEtran} 
\bibliography{arxiv} 

\end{document}